\documentclass[a4paper,fleqn,usenatbib]{mnras}

\usepackage[T1]{fontenc}
\usepackage{ae,aecompl}

\usepackage{graphicx}	
\usepackage{amsmath}	
\usepackage{amssymb}	
\usepackage{newtxtext,newtxmath}

\title[The open cluster NGC 225]{The young Galactic cluster NGC 225: binary stars' content and total mass estimate}

\author[Lidia Yalyalieva et al.]{
L. Yalyalieva,$^{1,2,3}$\thanks{E-mail: yalyalieva@yandex.ru}\
G. Carraro,$^{3}$\
E. Glushkova,$^{1,2}$\
U. Munari,$^{4}$\
and P. Ochner$^{3}$\\
$^{1}$Physics Department, Lomonosov Moscow State University, Leninskie Gory, Moscow, 119991, Russian Federation\\
$^{2}$Sternberg Astronomical Institute, Lomonosov Moscow State University, Universitetsky pr.13, Moscow, 119234, Russian Federation\\
$^{3}$Department of Physics and Astronomy, Padova University, Vicolo Osservatorio 3, I-35122, Padova, Italy\\
$^{4}$INAF National Institute of Astrophysics, Astronomical Observatory of Padova, 36012 Asiago, Italy\\
}
\date{Accepted XXX. Received YYY; in original form ZZZ}

\pubyear{}

\begin{document}
\label{firstpage}
\pagerange{\pageref{firstpage}--\pageref{lastpage}}
\maketitle

\begin{abstract}
Galactic star clusters are known to harbour a significant amount of binary stars, yet their 
role in the dynamical evolution of the cluster as a whole is not comprehensively understood. We investigated the influence of binary stars on the total mass estimate for the case of the moderately populated Galactic star cluster NGC 225. The analysis of multi-epoch radial velocities of the 29 brightest cluster members, obtained over two observational campaigns, in 1990-1991 and in 2019-2020, yields a value of binary fraction of $\alpha =0.52$ (15 stars
out of 29). Using theoretical isochrones and Monte Carlo simulations we found that the cluster mass increases at least 1.23 times when binaries are properly taken into account. By combining Gaia EDR3 photometric data with our spectroscopic observations, we derived estimates of NGC 225 fundamental  parameters foras follows: mean radial velocity $<V_r> = -9.8 \pm 0.7$ km\,s$^{-1}$, $\log(\tau)$=8.0-8.2 dex, distance $ D = 676 \pm 22$ pc, and colour excess $E(B-V) = 0.29 \pm 0.01$ mag.
 \end{abstract}




\section{Introduction}

Galactic star clusters (OCs) represent one of the scales in the hierarchy of star formation and their study does not loose relevance for many crucial reasons, as ampply discussed  in many recent studies \citep{CantatGaudin}. In fact, open clusters and associations are the birth place for most stars in the Galactic disc and hence their study makes it possible to trace the evolution not only of  individual star-forming regions but also of the disc as a whole \citep{Lada,Buckner}. However, our knowledge of these stellar systems is far from a complete, mostly because of the lack of good quality and statistically significant observational data.  For example, reliable mean radial velocities are available for a very small number of open star clusters.
This prevents accurate studies of star cluster internal kinematics, dynamics and chemistry. Besides, it makes it  impossible to re-construct orbits, which in turn are an essential information for  studying the kinematics of the disk of the Galaxy, to which the OCs subsystem belongs. High expectations in terms of obtaining accurate radial velocities are associated with the Gaia mission, especially with the most recent Data Release 3. However, the latest available Gaia Early Data Release 3 \citep{GaiaEDR3} inherited Gaia DR2 \citep{GaiaDR2} radial velocities, which in most cases are of insufficient precision for determining reliable mean cluster radial velocities. As an  example, in the catalog containing the mean radial velocities for 861 OCs according to Gaia DR2 data \citep{Soubiran}, for 60$\%$ of clusters the estimates of mean radial velocities are made using three or even fewer stars. In addition, with such small samples of stars having radial velocities  it seems challenging to take into account and then study the presence and relevance of binary stars in clusters. From theory, however, we know very well that they play a significant role in the dynamical evolution of the parent cluster: they boost mass segregation, they generate blue stragglers and other exotic system and, not less important, they are an important source of dynamical heating. Aong the same vein, their study is therefore important for the correct determination of the cluster mass \citep{Sheikhi} when using virial equilibrium.

The binary fraction (commonly indicate with $\alpha$) in open star clusters seems to be different than in  Galactic globular clusters and in the general Galactic field  \citep{Borodina}.

Photometry-based methods remain the most  popular way to estimate the binary fraction. The main approach of such studies is to unravel binaries according to their position above the main sequence of the parent cluster colour-magnitude diagram. \citet{Niu} used the synthetic colour-magnitude diagram method to derive $\alpha$ of 12 open clusters, which was found to range from 29 to 55 per\,cent for main sequence stars. \citet{Sollima} reported $\alpha$ from 35 to 70 per\,cent for 5 OC using  the same method. Also, basing on the stars's position in the colour-magnitude diagram, \citet{Khalaj} found binary fraction of 35 per\,cent for Praesepe (M44) and \citet{Sheikhi} reported $34 \pm 12$ per\,cent binary fraction for Alpha Persei open cluster. 

Some authors focused on more time-consuming spectroscopy-based methods to unravel binaries by studying radial velocities variations. \citet{Geller} summarised their observations of over 45 years for the old OC M67 and reported an overall $\alpha =34 \pm 3$ per\,cent and $70 \pm 17$ per\,cent for the cluster center. On the other hand, \citet{Banyard} estimated $\alpha$ to be $52 \pm8 $ per\,cent for the B-type stars of the OC NGC 6231.

The present study belongs to the latter group of studies. In fact, we aimed at obtaining radial velocities for all cluster members down to a certain magnitude and investigate the binary status of each of them. The open cluster NGC 225 in Cassiopea is a good target to this aim. It is a moderately populated cluster, located in an uncrowded region ($l \approx 122\degr.0$, $b \approx -1\degr.06$) of the sky. Investigations performed by different authors show  very scattered estimates of NGC 225 main physical parameters \citep{Bilir}, and in some cases even contradictory. \citet{Lattanzi} studied NGC 225 and reported estimates of its distance and age of $D= 525 \pm 73$ pc, $\tau = 120$ Myr, respectively. \citet{Subramaniam} claimed that the cluster is much younger, with the $\tau < 10$ Myr and distance $D = 575 \pm 120 $ pc. On the contrary, \citet{Bilir} reported a larger estimation of age, $\tau = 900 \pm 100$ Myr, with the same distance $D = 585 \pm 20 $ pc. \citet{Svolopoulos} found a distance $D = 630$ pc. Evaluation of color excess $E(B-V)$ varies from $0.151 \pm 0.047 $ \citep{Bilir} to $0.25 \pm 0.08$ mag \citep{Lattanzi} and 0.29 mag \citep{Svolopoulos,Subramaniam}. The most reliable estimation of mean proper motion in right ascension ($<\mu_{\alpha}*>$) and declination ($<\mu_\delta>$) were reported by \citet{CantatGaudin} based on Gaia DR2 \citep{GaiaDR2}: $<\mu_{\alpha}*>=-5.373$ \,mas\,year$^{-1}$ ,$<\mu_\delta>=-0.093$ \,mas\,year$^{-1}$; also the most likely distance was derived to be $D = 684.3$ pc.

Mean radial velocities estimations are contradictory too. \citet{Bilir} found a mean cluster radial velocity of $<V_r> = -8.3 \pm 5$ km\,s$^{-1}$ from eight stars in the cluster field. \citet{Conrad} instead reported $<V_r> = 28$ km\,s$^{-1}$, whilst \citet{Soubiran} used one star to determine mean radial velocity $<V_r> = -4.12 \pm 11.13$ km\,s$^{-1}$.

Our observational campaigns have therefore a twofold aim. On one side we attempt at amending this large discrepancies in the values of the cluster fundamental parameters. On the other side, we aim at determining its binary fraction and total mass.

To this purpose, the layout of this paper is as follows:  Section~\ref{sec:Memb} is dedicated to the selection of  cluster members, in Section~\ref{sec:Obs} we describe our observational data; in  Section~\ref{sec:PhysPar} we determine the binary fraction and assess how it affects the mass cluster estimation, while, finally, in Section~\ref{sec:Conclusion} we summarise our results.

\section{Membership probability}
\label{sec:Memb}
To  derive the binary status of each star down to a certain magnitude, the membership probability of the stars should be examined. This step is the very first step of this study and it was performed when only Gaia Data Release 2 \citep{GaiaDR2} was available. Therefore, we downloaded data from Gaia DR2 \citep{GaiaDR2} within $25 \arcmin$ radius centered on $\alpha = 0^h43^m31^s,  \delta = 61\degr 47\arcmin 43\arcsec $. Only data with parallax errors less than 20 per\,cent were taken into account. 
In order to distinguish probable cluster members from field stars we performed a clustering analysis using DBSCAN algorithm. DBSCAN - Density-Based Spatial Clustering of Applications with Noise - is a widely used method \citep{Gao,Castro-Ginard,Yalyalieva,Pasquato,Bhattacharya}, which separates all data points into a set of core points in the neighbourhood of each other, non-core points in the neighbourhood of core points, and noise. Two main parameters are needed to be fixed for the clustering analysis using DBSCAN: $\epsilon$ - the maximum distance between two points to label that one is in the neighbourhood of the other and {\it N} - the number of points in a neighbourhood of a point to label it as a core point.  The algorithm was implemented in Python language using the library \textsc{scikit-learn} \citep{scikit-learn}. The clustering was performed in a 3-dimensional space using parallaxes and the two components of proper motion. We avoid using positional coordinates because NGC 225 is quite sparse. Before proceeding with the clustering, the coordinates were scaled to unit dispersion and a  principal component analysis was performed to exclude possible dependencies between coordinates.

\begin{figure*}
	\includegraphics[width=\textwidth]{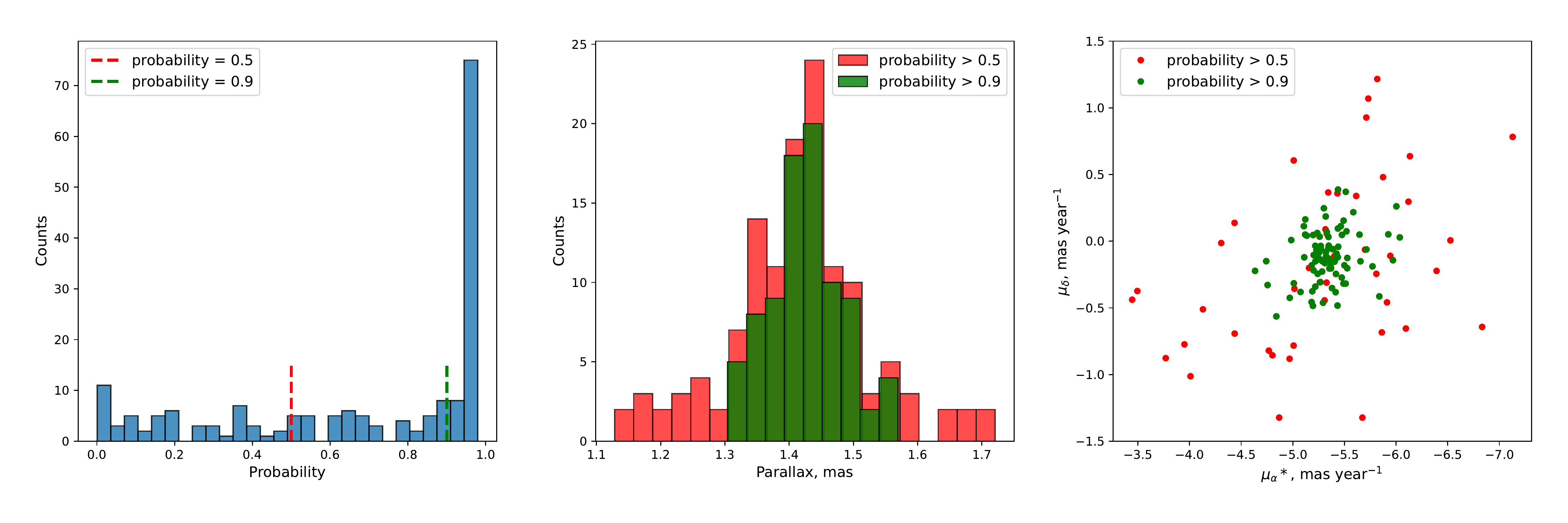}
	\caption{Left panel: distribution of membership probability; probability of all stars ever identified as cluster members are shown. Middle panel: parallax distribution for stars with membership more than 50 per cent (red) and 90 per cent (green). Right panel: proper motion vector-point diagram for stars with membership more than 50 per\,cent (red) and 90 per\,cent (green).}
	\label{fig:Prob}
\end{figure*}

Since it is usually rather complicated to find the best solution for $\epsilon$ and {\it N} values, we decided to adopt a statistical approach.\\ 
As a first step, we examined the vector-point diagram and the parallax distribution of the cluster stars. The bulk proper motion of stars of NGC 225 differs from the mean proper motion of foreground stars significantly. In fact, cluster stars crowd in a clump in vector-point diagram which is visible even by eye.
This clump is centred on the values reported by \citet{CantatGaudin} for the cluster proper motion components: $\mu_{\alpha}* \approx -5.4$ \,mas\,year$^{-1}$,  $\mu_\delta \approx -0.1$ \,mas\,year$^{-1}$, where $\mu_{\alpha}*=\mu_{\alpha}~cos\delta$ and $\mu_\delta$.  The stars with proper motion around this values have a mean parallax $\pi \approx 1.4$ mas. \\
As a second step, we constructed a grid of clustering parameters $\epsilon$ and {\it N}, extracting them from a wide interval ($\epsilon$ $= 0.01-0.99$, {\it N} $= 1-150$), and then we analysed the outcome of the clustering. Among all clustering solutions we selected those variants which produced only two clusters (groups): the group with mean parameters around the values found in the first step (group I), and the field stars group (group II). We found that the $n=1310$ clustering met this requirements. The number of stars identified as belonging to group I is 183. In a final third step we estimated the probability for a star to belong to NGC 225 by dividing the number of times when a star was related to the group I by the number of all possible clusterings $n$. The resulting distribution of the membership probability is shown in the Fig.~\ref{fig:Prob}. According to their probability, we compiled two lists of target stars: stars with probabilities $p > 50$ per\,cent (List$_{50}$), and stars with $p > 90$ per\,cent (List$_{90}$) (Fig.~\ref{fig:Prob}). Of the 183 stars in group I, 129 have a probability $\geq$0.5 and of these 85 are positioned above the 0.9 threshold. List$_{50}$ is presented in appendix (Table~\ref{tab:List50}), where stars are sorted by $G_{Gaia}$ magnitude and labeled from s001 for the brightest stars up to  s129 for the faintest one. In the following we adopted stars with membership probability $> 50$ per\,cent (that is from List$_{50}$), using stars from List$_{90}$ with membership probability $> 90$ for the most reliable estimations or as upper/lower limits. 

We compared results of our membership extraction with the list of members identified by \citet{CantatGaudin}. \citet{CantatGaudin} membership list includes 66 stars which lie in the same coordinates' space within $\approx 25 \arcmin$ in radius. We found that all those stars are present in our List$_{50}$ and 62 of them are also present in List$_{90}$.

As a result, following the method we described above, we were able to find the probability for a star to be a cluster member  avoiding to search for the exact values of $\epsilon$ and {\it N}.

In Table~\ref{tab:MeanParam} the mean values of the parallaxes and proper motions of the member stars are listed. As it may be noticed, the mean values of these two groups are quite similar. One could get the impression that the parallax dispersion is too large (see Fig.~\ref{fig:Prob}, middle panel). The maximum difference between the parallax of the star from List$_{50}$ and $<\pi>$ is $\Delta_{\pi} = 0.3$ mas, which translates into $\Delta_{\pi}/<\pi>^2 = 149$ pc, and a rough estimation for dispersion is $\sigma_{<\pi>}/<\pi>^2 = 50 $ pc. Initially, we adopted data with parallax errors not larger than 20 per\,cent, that means, again roughly, $0.2<\pi>/<\pi>^2 = 141$ pc. This is 3 times larger than the estimated dispersion and approximately equal to the maximum difference between star parallaxes and $<\pi>$. We cam therefore conclude that the results of clustering and membership assignment are quite reasonable within the limitations of the input data.

\begin{table}
	\centering
	\caption{Summary of mean parameters: mean parallax and its dispersion, mean components of the proper motion in right ascension and declination and corresponding dispersion.}
	\label{tab:MeanParam}
	\begin{tabular}{lcc}
Parameter / membership list                   & List$_{50}$       & List$_{90}$       \\
\hline
$<\pi>$, mas                                    & $1.42 \pm 0.01$   & $1.43 \pm 0.01$    \\
$\sigma_{<\pi>}$, mas\,year$^{-1}$            & $0.10 \pm 0.01$   & $0.06 \pm 0.01$    \\
<$\mu_{\alpha}*$>, mas\,year$^{-1}$           & $-5.33 \pm 0.02$  & $ -5.34 \pm 0.02$  \\  
$\sigma_{<\mu_{\alpha}*>}$, mas\,year$^{-1}$  & $0.46 \pm 0.01$   & $ 0.25 \pm 0.01$   \\
<$\mu_\delta$>, mas\,year$^{-1}$              & $-0.13 \pm 0.01$  & $ -0.11 \pm 0.01$  \\ 
$\sigma_{<\mu_{\delta}>}$, mas\,year$^{-1}$    & $0.37 \pm 0.01$   & $0.20 \pm 0.02$    \\
\hline
	\end{tabular}
\end{table}

It is important to note, finally, that the 30 brightest stars out of 183 (the whole number of stars ever labeled as group I stars) turned out to have membership probability larger than 50 per\, cent. Therefore, we are confident we are not going to miss any probable members among the first 30 brightest stars, which corresponds to $G_{Gaia}\leq 13^m.4$.

\subsection{Differences with Gaia EDR3 data}
As we performed membership segregation algorithm before Gaia Early Data Release 3 \citep{GaiaEDR3}, we based our results on Gaia Data Release 2 \citep {GaiaDR2}. To compare our results with the improved data from Gaia EDR3, we cross-correlated stars with membership probability > 50 per\,cent with Gaia EDR3 data and examined differences between parallaxes and proper motions. For most of the stars from List$_{50}$ the relative differences between parallaxes from Gaia DR2 and Gaia EDR3 was less than 10 per\,cent, and only one star (s051) exhibited a surprisingly significant smaller parallaxes in Gaia EDR3 than in Gaia DR2 data (difference $(\pi_{DR2}-\pi_{EDR3})/\pi_{DR2} \approx 0.45$ dex). The differences between each proper motion component for most of the stars were within 0.5 mas year$^{-1}$, but for the star s051 the difference in $\mu_{\delta}$ component was two times larger - about 1 mas year$^{-1}$'. We therefore decided to remove this star  from our sample. This change altered only $List_{50}$ and turned the number of the members down to 128.

\section{Observations and data reduction}
\label{sec:Obs}
In this study we used the results of two runs of the observations. Both of them were performed at the Asiago Astrophysical Observatory, Asiago, Italy. 

The first series of observations were carried out in 1990-1991 using the 1.8m Copernico telescope equipped with a Boller\,\&\,Chivens + CCD spectrograph. The CCD was a coated Thompson TH7882 and the recorded ranges for the 600 and 1200 ln/mm gratings were $\lambda=$ 3790-4910\,\AA \, and 3850-4400\,\AA, respectively. 

The second run was performed in 2019-2020. Observations were conducted with the 122cm Galileo telescope equipped with the Boller\,\&\,Chivens + CCD spectrograph with the 1200 ln/mm grating and range $\lambda=$ 3820-5035\,\AA. Besides,  several Echelle spectra ($\lambda=$ 3470 - 7360 \,\AA) were acquired using the 182cm Copernico telescope.

The reduction of the CCD frames has been done using the IHAP and the IRAF packages. In both series of observations flat fields have been regularly secured for each observing run, and calibration spectra (Fe/Ar lamp) were recorded before each observation of the target star.

Radial velocities of these two series of observations were extracted in different manner.
As for the runs made in 1990-1991, radial velocities have been derived for all target stars with cross correlation techniques using the routines implemented in the ESO-IHAP software package. The procedure we followed is described in detail in \citet{Munari92}. Briefly, it consists of an iterative process which leads to the definition of a template mean cluster for each observing run which is essentially composed by the normalised and added spectra of all non-binary cluster members. The definition of this template mean cluster has been greatly facilitated by the narrow range of spectral types and projected rotational velocities spanned by NGC 225 target stars. The typical error of the single observation is $\approx 3$ km\,sec$^{-1}$.

During runs in 2019-2020, to obtain radial velocities from the spectra we used a Fourier cross-correlation technique implemented in IRAF environment. As a template spectra we used synthetic spectra from \citet{Munari2005} with atmospheric parameters appropriate for each spectral type from \citet{Landolt1982}.
Spectral classifications were derived according to \citet{Gray}. Synthetic template spectra for each processing star were then chosen according to the derived spectral classes.

All the data discussed in this study is  made public available at the Centre de Donn{\'e}es Stellaire data-base.

\section{Physical parameters}
\label{sec:PhysPar}

\subsection{Binary fraction and mean radial velocity}
\label{sec:Spectroscopy}

The results of the radial velocities extraction is listed in Table~\ref{tab:RV}, while spectral classification is reported in Table~\ref{tab:ST}.

\begin{table*}
	\centering
	\caption{Radial velocities of the 29 brightest stars of NGC 225. Symbol "*" near the date of the observation refers to Echelle observations.}
	\label{tab:RV}
	\begin{tabular}{lcc | lcc | lcc | lcc}
		\hline
		star & date,     & $V_r$,   &  star & date,     & $V_r$,   &    star & date,     & $V_r$,   &  star & date,     & $V_r$,         \\
		 & dd.mm.yyyy &  km\,s$^{-1}$ & & dd.mm.yyyy &  km\,s$^{-1}$  & & dd.mm.yyyy &  km\,s$^{-1}$ & & dd.mm.yyyy &  km\,s$^{-1}$\\
		\hline
		s001 & 09.01.90 & -10  &  s004 & 09.01.90 & -5     &   s008 & 04.12.19 & -9.8   &  s014 & 23.01.20 & -11.8   \\
		& 17.01.90 & -12      &   & 17.01.90 & -10        &    & 22.01.20 & -10.3      &   & 20.11.20 & -10.2       \\  \cline{10-12}
		& 18.02.90 & -5       &   & 18.02.90 & -8         &    & 23.01.20 & -9.0       &  s015 & 19.02.90 & 22      \\
		& 19.02.90 & -12      &   & 19.02.90 & -15        &    & 20.11.20 & -7.9       &   & 28.09.90 & -3          \\
		& 28.09.90 & -2       &   & 28.09.90 & -9         &    & 07.02.20* & -11.1     &   & 02.10.90 & -15         \\ \cline{7-9}
		& 02.10.90 & -11      &   & 14.10.90 & 0          &   s009 & 09.01.90 & -9     &   & 14.10.90 & -8          \\  
		& 14.10.90 & -8       &   & 08.01.91 & -9         &    & 18.02.90 & -11        &   & 06.11.90 & 25          \\
		& 06.11.90 & -6       &   & 24.01.91 & -8         &    & 19.02.90 & -11        &   & 08.01.91 & 4           \\
		& 08.01.91 & -9       &   & 25.01.91 & -9         &    & 28.09.90 & -10        &   & 26.01.91 & -10         \\
		& 24.01.91 & -12      &   & 26.01.91 & -9         &    & 02.10.90 & -7         &   & 04.12.19 & -9.6        \\
		& 25.01.91 & -6       &   & 04.12.19 & -10.9      &    & 14.10.90 & -5         &   & 22.01.20 & -9.7        \\
		& 26.01.91 & -12      &   & 22.01.20 & -9.3       &    & 06.11.90 & -2         &   & 20.11.20 & -9.6        \\  \cline{10-12}
		& 04.12.19 & -9.7     &   & 19.11.20 & -9.2       &    & 08.01.91 & -5         &  s016 & 12.01.20 & -15.8   \\ \cline{4-6}
		& 11.01.20 & -10.4    &  s005 & 09.01.90 & -5     &    & 24.01.91 & -3         &   & 20.11.20 & -12.6       \\ \cline{10-12}
		& 11.01.20 & -12.8    &   & 18.02.90 & 10         &    & 26.01.91 & -11        &  s017 & 04.12.19 & -13.9   \\
		& 19.11.20 & -10.3    &   & 19.02.90 & -10        &    & 04.12.19 & -9.2       &   & 23.01.20 & -13.5       \\ \cline{1-3}
		s002 & 09.01.90 & -9   &   & 28.09.90 & -19        &    & 22.01.20 & -9.9       &   & 21.11.20 & -12.9       \\ \cline{10-12}
		& 17.01.90 & -8       &   & 02.10.90 & 4          &    & 20.11.20 & -8.4       &  s018 & 12.01.20 & -10.1   \\ \cline{7-9}
		& 18.02.90 & -11      &   & 14.10.90 & -16        &   s010 & 04.12.19 & -9.4   &   & 21.11.20 & -11.8       \\ \cline{10-12}
		& 19.02.90 & -8       &   & 06.11.90 & -28        &    & 22.01.20 & -11.3      &  s019 & 04.12.19 & -10.9   \\
		& 28.09.90 & -1       &   & 08.01.91 & -27        &    & 23.01.20 & -11.7      &   & 23.01.20 & -11.9       \\
		& 02.10.90 & -7       &   & 24.01.91 & -28        &    & 20.11.20 & -9.3       &   & 21.11.20 & -10.4       \\ \cline{10-12}
		& 14.10.90 & -5       &   & 26.01.91 & -10        &    & 07.02.20* & -11.4     &  s020 & 04.12.19 & -10.8   \\ \cline{7-9}
		& 06.11.90 & -12      &   & 04.12.19 & -8.3       &   s011 & 28.09.90 & -1     &   & 21.11.20 & -10.3       \\ \cline{10-12}
		& 08.01.91 & -13      &   & 22.01.20 & -11.6      &    & 02.10.90 & -8         &  s021 & 23.01.20 & -23.1   \\
		& 24.01.91 & -15      &   & 20.11.20 & -8.2       &    & 14.10.90 & -10        &   & 21.11.20 & -14.5       \\ \cline{10-12} \cline{4-6}
		& 25.01.91 & -13      &  s006 & 19.02.90 & -6     &    & 06.11.90 & 1          &  s022 & 11.01.20 & -9.9    \\
		& 26.01.91 & -8       &   & 28.09.90 & -10        &    & 08.01.91 & -20        &   & 21.11.20 & -10.6       \\ \cline{10-12}
		& 04.12.19 & -8.2     &   & 02.10.90 & -7         &    & 26.01.91 & -1         &  s023 & 11.01.20 & -15.8   \\
		& 22.01.20 & -7.6     &   & 14.10.90 & -8         &    & 04.12.19 & -7.1       &   & 23.01.20 & -15.6       \\
		& 19.11.20 & -8.7     &   & 06.11.90 & -12        &    & 23.01.20 & -12.0      &   & 21.11.20 & -12.7       \\ \cline{10-12} \cline{1-3}
		s003 & 09.01.90 & -5   &   & 08.01.91 & -6         &    & 20.11.20 & -12.7      &  s024 & 04.12.19 & -9.8    \\ \cline{7-9}
		& 17.01.90 & -3       &   & 24.01.91 & -12        &   s012 & 04.12.19 & -9.9   &   & 21.11.20 & -10.4       \\ \cline{10-12}
		& 18.02.90 & -31      &   & 04.12.19 & -9.8       &    & 23.01.20 & -11.0      &  s025 & 12.01.20 & -17.1   \\
		& 28.09.90 & -16      &   & 11.01.20 & -9.4       &    & 20.11.20 & -10.3      &   & 21.11.20 & -13.5       \\ \cline{10-12}
		& 02.10.90 & -3       &   & 22.01.20 & -9.5       &    & 07.02.20* & -8.3      &  s026 & 11.01.20 & -15.5   \\ \cline{7-9}
		& 14.10.90 & 8        &   & 20.11.20 & -8.6       &   s013 & 09.01.90 & -15    &   & 21.11.20 & -22.3       \\ \cline{10-12}  \cline{4-6}
		& 06.11.90 & -14      &  s007 & 09.01.90 & -9     &    & 28.09.90 & -9         &  s027 & 11.01.20 & -19.5   \\
		& 08.01.91 & -32      &   & 28.09.90 & 1          &    & 02.10.90 & 13         &   & 22.11.20 & 1.2         \\ \cline{10-12}
		& 24.01.91 & -6       &   & 02.10.90 & -3         &    & 14.10.90 & 10         &  s028 & 12.01.20 & -6.3    \\
		& 25.01.91 & -2       &   & 14.10.90 & -9         &    & 06.11.90 & 26         &   & 22.11.20 & -5.6        \\
		& 26.01.91 & 20       &   & 06.11.90 & -12        &    & 08.01.91 & 30         &   & 22.11.20 & -5.1        \\ \cline{10-12}
		& 04.12.19 & -4.8     &   & 08.01.91 & -6         &    & 24.01.91 & -12        &  s029 & 12.01.20 & -2.9    \\
		& 22.01.20 & -2.0     &   & 24.01.91 & 1          &    & 26.01.91 & 32         &   & 22.11.20 & -42.6       \\
		& 19.11.20 & -8.8     &   & 26.01.91 & -6         &    & 04.12.19 & -11.4      &   & 22.11.20 & -41.9       \\
		&          &          &   & 04.12.19 & -9.2       &    & 20.11.20 & -11.7      &   &          &             \\
		&          &          &   & 22.01.20 & -9.1       &    &          &            &   &          &             \\
		&          &          &   & 20.11.20 & -7.6       &    &          &            &   &          &             \\
		\hline
	\end{tabular}
\end{table*}

\begin{table}
	\centering
	\caption{Spectral classification}
	\label{tab:ST}
	\begin{tabular}{lcc} 
		\hline
	N      &  Spectral type  & Spectral type   \\
	&                 & \citet{Lattanzi}    \\
	\hline
	s001   &  B7             &  B6.5                   \\
	s002   &   B8            &  B8                     \\
	s003   &   B8            &  B9                     \\
	s004   &   B9            &  A0                     \\
	s005   &   B8III         &  B9                     \\
	s006   &   B9            &  A0                     \\
	s007   &   A1            &  A1                     \\
	s008   &  A1             &  -                      \\
	s009   &   A0            &  A0                     \\
	s010   &   A3            &  A9                     \\
	s011   &   A1            &  A0                     \\
	s012   &   A2            &  -                      \\
	s013   &   A1            &  A3                     \\
	s014   &   A2            &  -                      \\
	s015   &   A0III         &  A2                  \\       
	s016   &  A3            &  A5                 \\
	s017   &  A5            &  -                  \\
	s018   &  F2            &  A7                 \\
	s019   &   F0           &  -                  \\
	s020   &  F0            &  A7                 \\
	s021   &   A3           &                     \\
	s022   &   F0           &  F0                 \\
	s023   &   F2           &  -                  \\
	s024   &   F6           &  F6                 \\
	s025   &  F7            &                     \\
	s026   &   F5           &  F4                 \\
	s027   &  F5            &  F3                 \\
	s028   &  F5            &  F7                 \\
	s029   &  F8            &  -                  \\
	
		\hline
	\end{tabular}
\end{table}

\noindent
To single out binary stars, we used the following criteria:

\begin{description}
	\item (a) If a star had more than 3 observations, we used the Pearson's chi-squared test with 95 per\,cent significance level. This was applied to 11 stars.
	
	\item (b) If a star had only 3 or 2 observations, we calculated mean radial velocity ($V_r$) and dispersion ($\sigma$) and then compared them with the cluster mean radial velocity ($V_{Cl}$) and its dispersion ($\sigma_{Cl}$). If $V_r$ differs from $V_{Cl}$ by $3\times \sigma_{Cl}$ or $\sigma > 3\times \sigma_{Cl}$, the star was assumed to be a binary.
\end{description}
	
	The mean cluster radial velocity and its dispersion were estimated by means of an iterative process. We used only {\it bona fide} single stars. At the first step, we used only stars marked as single by (a) criterion. Then stars marked as single by criterion (b)  were added and values of $V_{Cl}$ and $\sigma_{Cl}$ were calculated again. Iterations were repeated until convergence. The resulting values of mean cluster radial velocity and dispersion are: $V_{Cl} = -9.8 \pm 0.7$ km\,s$^{-1}$, $\sigma_{Cl} = 1.0 \pm 0.1$ km\,s$^{-1}$.
	
	Filtering through (a) and (b) criteria we counted 15 binary stars among 29 stars belonging to the List$_{50}$ or 12 stars among 22 stars belonging to the List$_{90}$, which implies a fraction of binary stars $\alpha=0.52$ and $\alpha=0.55$, respectively (see Table~\ref{tab:binarity}).
	
\begin{table}
	\centering
	\caption{The number of obtained spectra, mean radial velocity and its dispersion, binary status, and membership probability for the 29 brightest stars in NGC 225.}
	\label{tab:binarity}
	\begin{tabular}{lccccc} 
		\hline
		star & number of & $<V_r>$, & $\sigma_{V_r}$, & binarity & membership\\
		 & spectra & km\,s$^{-1}$ & km\,s$^{-1}$ &  & probability \\
		\hline
	s001 & 16 & -9.3 & 3.1 & single     & 0.976  \\
	s002 & 15 & -9.0 & 3.5 & single & 0.775  \\
	s003 & 14 & -7.1 & 13.5 &binary & 0.958  \\
	s004 & 13 & -8.6 & 3.4 & single & 0.692  \\
	s005 & 13 & -12.1 & 11.6&binary & 0.862  \\
	s006 & 11 & -8.9 & 2.1 & single & 0.977  \\
	s007 & 11 & -6.3 & 4.3 & binary & 0.970  \\
	s008 & 5 & -9.6 & 1.2 & single  & 0.978  \\
	s009 & 13 & -7.8 & 3.1 & single & 0.977  \\
	s010 & 5 & -10.6 & 1.2 & single & 0.979  \\
	s011 & 9 & -7.9 & 6.8 & binary  & 0.979  \\
	s012 & 4 & -9.9 & 1.2 & single  & 0.976  \\
	s013 & 10 & 5.2 & 19.2 & binary & 0.852  \\
	s014 & 2 & -11.0 & 1.1 & single & 0.682  \\
	s015 & 10 & -1.4 & 14.1 &binary & 0.977  \\
	s016 & 2 & -14.2 & 2.3 & binary & 0.955  \\
	s017 & 3 & -13.4 & 0.5 & binary & 0.970  \\
	s018 & 2 & -10.9 & 1.2 & single & 0.845  \\
	s019 & 3 & -11.1 & 0.8 & single & 0.978  \\
	s020 & 2 & -10.6 & 0.3 & single & 0.979  \\
	s021 & 2 & -18.8 & 6.1 & binary & 0.638  \\
	s022 & 2 & -10.3 & 0.5 & single & 0.977  \\
	s023 & 3 & -14.7 & 1.7 & binary & 0.977  \\
	s024 & 2 & -10.1 & 0.4 & single & 0.975  \\
	s025 & 2 & -15.3 & 2.6 & binary & 0.959  \\
	s026 & 2 & -18.9 & 4.8 & binary & 0.977  \\
	s027 & 2 & -9.1 & 14.6 & binary & 0.977  \\
	s028 & 3 & -5.7 & 0.6 & binary  & 0.979  \\
	s029 & 3 & -29.1 & 22.7& binary & 0.977  \\
		\hline
	\end{tabular}
\end{table}

\subsection{Distance modulus and age}
\label{sec:Photometry}

In the next section we are going to investigate how the obtained value of binary fraction affects the mass estimations of the cluster. We used isochrones to extract from the magnitude of the each star the corresponding mass. To this purpose, we first of all calculated the distance module by fitting with theoretical isochrone  the position of member stars in the color-magnitude diagram (CMD). The fit was performed eye-balling the stars' distribution in the CMD.

\begin{figure*}
	\includegraphics[width=\textwidth]{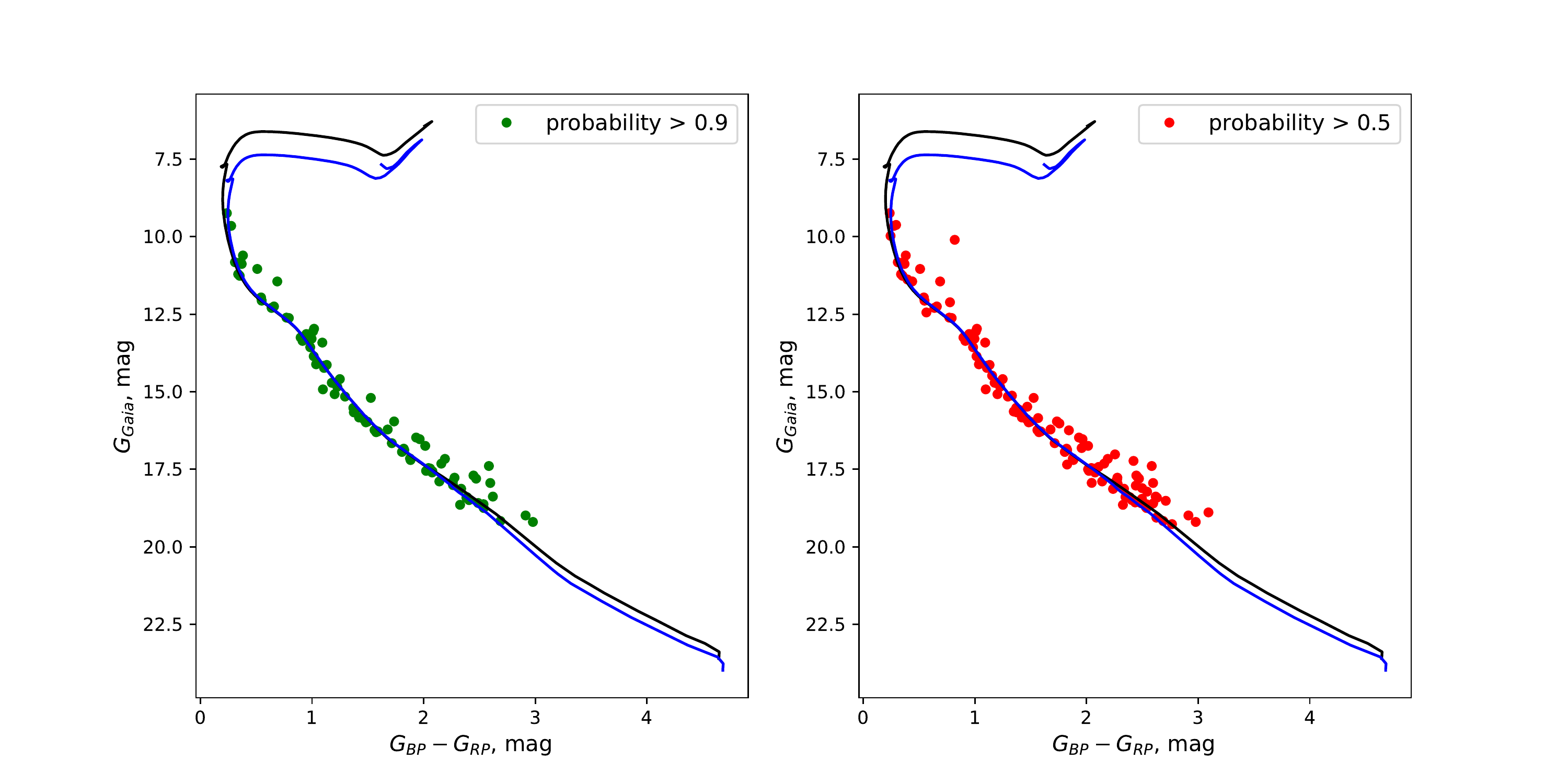}
	\caption{The color-magnitude diagram ($G_{Gaia}$,$G_{BP}-G_{RP}$). Black and blue lines - shifted isochrones with $\log{\tau}=8.0$ dex and $8.2$ dex correspondingly. Left panel represent stars with a membership probability greater than 90 per\,cent, right panel - with a membership probability greater than 90 per\,cent.}
	\label{fig:GaiaIsoch}
\end{figure*}

By fitting theoretical PARSEC + COLIBRI isochrones \citep{Bressan12} to Gaia EDR3 photometry (see Fig.~\ref{fig:GaiaIsoch}), we found the following parameters:

\noindent 
$(m-M)_{G}=9.90 \pm 0.06$ mag\\
$E(G_{BP}-G_{RP})=0.40 \pm 0.02$ mag,\\
$\log{\tau} = 8.0 - 8.2$ dex or 100-160 Myr,\\
 where $G_{BP}$ and $G_{RP}$ are Gaia blue and red pass-bands, see Fig.~\ref{fig:GaiaIsoch}.
 Using \citet{Cardelli} coefficients we obtained $E(B-V) = 0.29 \pm 0.01$ mag, $(m-M)_0=9.15 \pm 0.07$ mag and distance $D=676 \pm 22$ pc. 

We compared our photometric distance with data from \citet{BailerJones}, which are based on Gaia EDR3 data. Taking the median value of {\it rgeo} - median of the geometric distance, {\it b\_rgeo} - 16th percentile of the geometric distance, and {\it B\_rgeo} - 84th percentile of the geometric distance, we obtained 682 pc, 664 pc and 707 pc, respectively. The distance reported by \citet{CantatGaudin} is $D=684.3$ pc. Both values show therefore good agreement with our estimate.

On the other side, our age estimation is in good agreement with the one reported by \citet{Lattanzi}, namely about 120 Myr.

We derived the same value of the color excess $E(B-V)$ as in \citet{Svolopoulos} and \citet{Subramaniam}.
We compared this value with the data of 3-dimensional map of dust reddening \citet{Green}. We extracted data in the direction of each star from the $List_{50}$ for the distance of $D=676$ pc and, assuming $R_v=3.1$ and using coefficients from \citet{Schlafly} for conversion, we obtained a mean color excess of $E(B-V)=0.29$ mag.
Also the 3-dimensional medium map STILISM \citep{Capitanio} for the same distance $D=676$ pc yields a mean color excess $E(B-V)=0.31$ mag, in good agreement with the value we obtained.

All these results highlight the advantages to perform ishocrone fitting when a solid sample of member stars is available.

\subsection{Cluster mass}
Before calculating the mass of the cluster, it is mandatory to assess the photometric completeness of our sample.

\begin{figure}
	\includegraphics[width=\columnwidth]{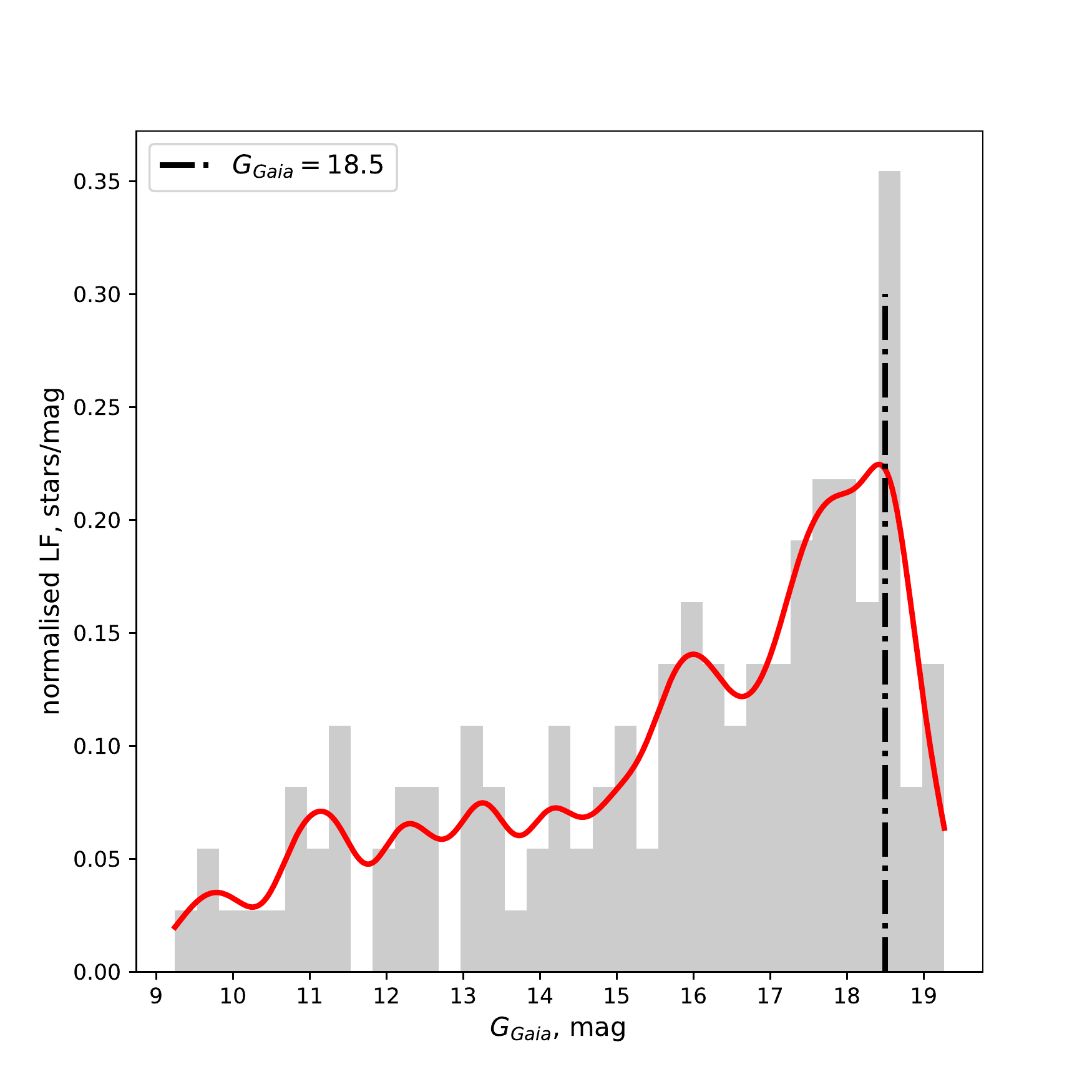}
	\caption{Normalised distribution of Gaia $G$ filter for stars with membership probability > 50 per\,cent. Dash-dotted line indicates $G_{Gaia}=18^m.5$.}
	\label{fig:normLF}
\end{figure}

 In \citet{Boubert} it is claimed that Gaia DR2 is complete down to $G = 18^m.9 - 21^m.3$. depending on the sky coordinates. It seems reasonable to assume that Gaia EDR3 limit is not worse than that.
In the Fig.~\ref{fig:normLF} the normalised distribution of Gaia EDR3 $G$ magnitude for stars from our List$_{50}$ (membership probability > 50 per\,cent) is shown. The red line is the output of the kernel density estimator applied to this data. We used Gaussian function as a kernel with a bandwidth $=0.3$. We chose this value under the same assumptions as  in \citet{Seleznev}. Below $G_{Gaia} \approx 18^m.5$ a sharp decrease is noticed which we interpreted as photometric  completeness limit. In the Fig.~\ref{fig:normLF} this limit is indicated with a dashed-dot line. The whole number of stars having $G_{Gaia} \leq 18^m.5$ is 110.

We then estimated the cluster mass by assuming two different values for the binary fraction $\alpha$.

\subsubsection{Binarity fraction $\alpha = 0$}
\label{BF0}
In the first case we assign a binary fraction $\alpha = 0$.
To estimate the cluster mass down to$ G_{Gaia} = 18^m.5$ we calculated first the absolute magnitude of each star in $G_{Gaia}$ using $(m-M)_G=9.90 \pm 0.06$ mag (see Section~\ref{sec:Photometry}). Then we find a transformation function from absolute magnitude to mass $M(G_{Gaia})$ by spline interpolating data from \citep{Bressan12} isochrones. By summing up the masses of all stars with visual $G_{Gaia}\lid 18^m.5$ we obtained the cluster mass $\mathcal{M}_0$. Adding and subtracting $0.06$ from $(m-M)_G$ we obtained the associated uncertainty. Hence, we get $\mathcal{M}_0 = 126.6 \pm 1.7$ M$_{\sun}$ for an isochrone with the $\log{\tau} = 8.0$,  and $\mathcal{M}_0 = 125.3 \pm 1.6 $ M$_{\sun}$ for an isochrone with the $\log{\tau} = 8.2$. The mass that corresponds to $G_{Gaia}=18^m.5$ is $\mathcal{M}_{lim}=0.54$ M$_{\sun}$ for $\log{\tau} = 8.0$ dex and $\mathcal{M}_{lim}=0.53$ M$_{\sun}$ for $\log{\tau} = 8.2$ dex.

\begin{figure*}
	\includegraphics[width=\textwidth]{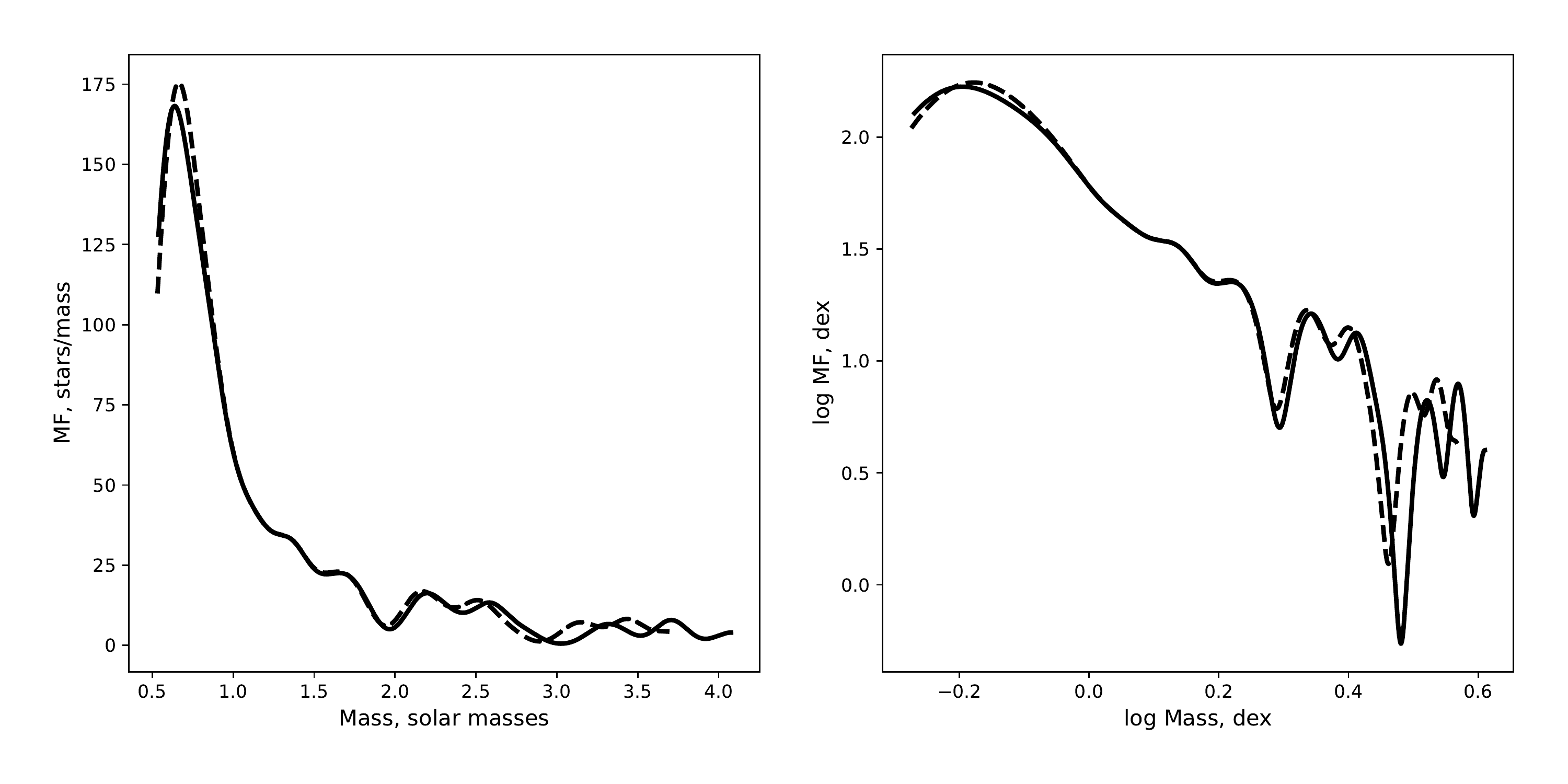}
	\caption{Left panel: mass function, obtained with Gaia isochrone $\log{\tau} = 8.0$ (black line) and $\log{\tau} = 8.2$ (black dashed line). Right panel:  the same, but in logarithmic scale.}
	\label{fig:MF}
\end{figure*}

Knowing the distribution of the mass down to $G_{Gaia} = 18^m.5$, and applying kernel density estimator with Gaussian kernel and bandwidth parameter $=0.2$, we constructed the cluster mass function (MF) (see Fig.~\ref{fig:MF}).
Then, by applying a least-square fit to the logarithmic MF in mass range $0.8$ M$_{\sun}< \mathcal{M} <1.9$ M$_{\sun}$, we found the slope coefficient $a$. Taking into account the uncertainty in $(m-M)_G$ and recalculating the  mass of each star according to it for both $\log{\tau} = 8.0$ and $\log{\tau} = 8.2$, we finally obtained $a = -2.53 \pm 0.02$.

We remind the reader that the standard Salpeter IMF has a slope of $a = -2.35$ while  \citet{Kroupa} proposed $a=-1.3 \pm 0.5$ for stars with mass $0.08$ M$_{\sun} < \mathcal{M}< 0.5$ M$_{\sun}$, $a= -2.3 \pm 0.3$ for $ 0.5$ M$_{\sun} < \mathcal{M} < 1$ M$_{\sun}$ and $a = -2.3 \pm 0.7$ for mass range $\mathcal{M} > 1$ M$_{\sun}$.

\subsubsection{Binary fraction $\alpha = 0.52$}

A binary fraction $\alpha = 15/29 \approx 0.517$ is the result which we obtained in Section~\ref{sec:Spectroscopy} for the stars with membership probability larger than 50 per\,cent.

We used the same system of equations that were recently used in \citet{Borodina}, namely:

\begin{equation}
\begin{cases}
L=L_1+L_2,\\
L_1=L(\mathcal{M}_1),\\
L_2=L(\mathcal{M}_2),\\
q=\mathcal{M}_2/\mathcal{M}_1,\\
\mathcal{M}=\mathcal{M}_1+\mathcal{M}_2,\\
\end{cases}\,.
\label{eq:L_system}
\end{equation}

where $L$ and $\mathcal{M}$ are the total luminosity and total mass of the binary system, $L_1$, $L_2$ and $\mathcal{M}_1$,$\mathcal{M}_2$ are the luminosity and mass of the components, $\mathcal{M}_1\gid \mathcal{M}_2$ so $0 < q \lid 1$.

We also adopted the following equation from \citet{Eker15}:
\begin{equation}
log{L}=-0.705\times(log{\mathcal{M}})^2+4.655\times(log{\mathcal{M}})-0.025
\end{equation}
Using isochrones from \citep{Bressan12} and the distance modulus $(m-M)_{G}$ previously derived, we transformed Gaia $G$ magnitudes into luminosities $L$. To find a solution for the equation~(\ref{eq:L_system}) we used optimisation algorithm from the Python library \textsc{SciPy} \citep{SciPy} and minimised the following function using a truncated Newton minimisation algorithm:
\begin{equation}
y={\lvert}L-10^{L(\mathcal{M}_1)}-10^{L(q\mathcal{M}_1)}{\rvert}
\end{equation}

Under the assumption that the binary fraction does not depend on the magnitude interval (hence mass), we may consider $\alpha$ as a probability that a randomly chosen star turned out to be a binary. 
We can therefore assign to each star a value $\beta$ randomly extracted in the interval $[0,1]$.  When $\beta$ turns out to be $\lid \alpha$ we treat that star as a binary and applied minimisation algorithm to find $\mathcal{M}_1$ and then $\mathcal{M}=\mathcal{M}_1\times(1+q)$. In the opposite case, we assign to it mass of a single stars as described in Section~\ref{BF0}. Repeating that procedure for 1000 runs, we found mean values of the total cluster mass estimation $\mathcal{M}_{\alpha}$. Dividing this value by the mass $\mathcal{M}_0$, as calculated in Section~\ref{BF0}, we derived the cluster mass increment $\nu$ induced by binaries.

We tested two cases as far as the $q$ distribution is concerned: $q=1$ and a flat $q$ distribution. In the latter case, after each $\mathcal{M}_1$ calculation we checked whether the value $\mathcal{M}_1\times q$ is greater than $0.08$ M$_{\sun}$. If this condition was not met, the $q$ value was chosen again.
The calculations were done for all the stars with $G_{Gaia}\lid 18^m.5$. Uncertainties in masses were calculated by taking into account uncertainties in distance moduli. The results are listed in Table~\ref{tab:Masses}. We emphasize that the difference between the two age values ($\log{\tau}=8.0$ dex and $\log{\tau}=8.2$ dex) is marginal and well within the  uncertainties.\\

\begin{table}
	\centering
	\caption{Total cluster mass, its dispersion, and cluster mass increment $\nu$.}
	\label{tab:Masses}
	\begin{tabular}{lcccc} 
		\hline
distribution                     & $log{\tau}$,  & $\mathcal{M}_\alpha$,       &   $\sigma_{\mathcal{M}_\alpha}$,&   $\nu$,\\
& dex & M$_{\sun}$ & M$_{\sun}$ &dex\\\hline                                                                                    
q=1                       & 8.0         & $170.5 \pm 2.2 $  &   4.5                 &   1.35  \\
& 8.2         & $168.6 \pm 2.3 $  &   4.8                 &   1.35  \\     
flat                      & 8.0         & $155.2 \pm 2.0 $  &   3.7                 &   1.23  \\
distribution              & 8.2	        & $153.6 \pm 1.9 $  &   3.6                 &  1.23	  \\

		\hline
	\end{tabular}
\end{table}

\noindent
For the case of $q=1$ we calculated the total mass for the 29 stars for which we had derived the binary status using our spectroscopic criteria. As we had already found binary stars among them, we could calculate their mass in accordance with their binary status and could compare the results obtained for the case of randomly assigned binary stars. The difference did not exceed $4$ M$_{\sun}$ for all cases, or less than 5 per\,cent. 

The obtained cluster mass increments $\nu$ are in good agreement with the values obtained by \citet{Borodina}, where for $\alpha \approx 0.5$ $\nu$ was found to be about 1.2 for "realistic" distributions of q, including a flat distribution, and $\nu \approx 1.35$ for q=1.

In this study we considered only binaries, and not multiple systems. \citet{Borodina21} demonstrates that taking into account triple and quadruple systems inflates $\nu$ significantly. Therefore,  our estimations, especially for flat $q$ distribution, should be considered as a lower limit.

\section{Conclusions}
\label{sec:Conclusion}
In this work we studied the open star cluster NGC 225 and investigated how the presence of binary stars among cluster members affects the cluster total mass measure. The main results of our study  can be summarised as follows:
\begin{description}
    \item (1) By using Gaia DR2 data and applying a clustering analysis to stars' proper motions and parallaxes, we obtained a list of probable cluster members. All  29 brightest stars from this list have membership probability larger than 50 per\,cent. By retaining stars with this probability value or larger, we guarantee that the list of bright stars is complete.
    \item (2) We employed spectroscopic observations, collected in 1990-1991 and 2019-2020, and derived radial velocities for the brightest 29 cluster members down to G$=13^m.4$ in the $G$ Gaia EDR3 filter. The binary fraction among this 29 stars turned out to be $\alpha = 0.52$.  The mean cluster radial velocity we derived is $<V_r>= -9.8 \pm 0.7$ km\,s$^{-1}$.
    \item (3) The fit with theoretical isochrones to Gaia EDR3 photometric data yields a cluster distance $ D = 676 \pm 22$ pc, an age $\log{\tau}=8.0$-$8.2$ dex, and a colour excess $E(B-V) = 0.29 \pm 0.01$ mag.
    \item (4) The cluster mass calculation was performed down to $G_{Gaia}=18^m.5$ for two binary fraction values: $\alpha =0$ and $\alpha=0.52$ (15 stars out of 29). We used isochrones for mass assignment for single stars, and in the case of $\alpha=0.52$ we applied Monte Carlo simulations. We tested two types of binary mass distribution: 'flat' and the case of equal mass components. The results of total mass estimations down to $G=18^m.5$ Gaia EDR3 magnitude varies from $\mathcal{M}_0 =125.3 \pm 1.7$ M$_{\sun}$ for the case $\alpha =0$ to $\mathcal{M}_{\alpha} =155.2 \pm 2.0$ M$_{\sun}$ ('flat' mass distribution) and $\mathcal{M}_{\alpha} =170.5 \pm 2.2$ M$_{\sun}$ (equal mass components) for the case $\alpha =0.52$. The comparison shows that the cluster mass increases between 1.23 and 1.35 times if the binary fraction is taken into account.
\end{description}

\section{Data Availability}
The data underlying this article are available in the article and in its online supplementary material.

\section*{Acknowledgements}
The reported study was funded by RFBR according to the research project number 20-32-90124. L. Yalyalieva acknowledges the financial support of the bilateral agreement between Lomonosov Moscow State University and Padova University, that allowed her to spend a period in Padova, where this work was initiated. The work of  G. Carraro has been supported by Padova University grant BIRD191235/19: {\it Internal dynamics of Galactic star clusters in the Gaia era: binaries, blue stragglers, and their effect in estimating dynamical masses}.\\
\noindent
This work made use of data from the European Space Agency (ESA) mission
{\it Gaia} (\url{https://www.cosmos.esa.int/gaia}), processed by the {\it Gaia}
Data Processing and Analysis Consortium (DPAC,
\url{https://www.cosmos.esa.int/web/gaia/dpac/consortium}). Funding for the DPAC
has been provided by national institutions, in particular the institutions
participating in the {\it Gaia} Multilateral Agreement.




\bibliographystyle{mnras}
\bibliography{bibtex} 




\appendix

\section{List of cluster members}

\begin{table*}
	\centering
	\caption{List of cluster members with membership probability > 50 per\,cent. $G_{Gaia}$ magnitude, proper motion and parallaxes are from Gaia DR2. Symbol "*" refers to rejected star.}
	\label{tab:List50}
	\begin{tabular}{lcccccccc} 
		\hline
Star & Right Ascention, & Declination, & $G_{Gaia}$,& Parallax, & $ \mu_{\alpha}~cos\delta$, & $\mu_{\delta},$  & Probability & Probability from\\
& h:m:s           & d:m:s       & mag & mas & mas\,year$^{-1}$ & mas\,year$^{-1}$ & & \citet{CantatGaudin}    \\
\hline

	s001 & 00:44:40.81 & +61:48:43.3 & 9.2 & 1.42 & -5.01 & -0.32 & 0.98 & 0.2\\
	s002 & 00:44:30.67 & +61:46:49.9 & 9.6 & 1.4 & -6.13 & 0.64 & 0.77 & -\\
	s003 & 00:44:40.46 & +61:54:01.8 & 9.6 & 1.32 & -5.42 & -0.13 & 0.96 & -\\
	s004 & 00:44:46.41 & +61:52:31.4 & 10.0 & 1.44 & -6.83 & -0.64 & 0.69 & -\\
	s005 & 00:43:26.58 & +61:45:55.7 & 10.1 & 1.36 & -5.87 & 0.48 & 0.86 & 0.5\\
	s006 & 00:43:51.06 & +61:50:08.3 & 10.6 & 1.41 & -5.24 & 0.06 & 0.98 & 0.8\\
	s007 & 00:43:51.47 & +61:47:13.5 & 10.8 & 1.35 & -5.12 & 0.05 & 0.97 & 0.5\\
	s008 & 00:43:21.88 & +61:27:10.7 & 10.8 & 1.39 & -5.24 & -0.12 & 0.98 & 0.5\\
	s009 & 00:43:28.88 & +61:48:04.0 & 10.9 & 1.44 & -5.24 & -0.24 & 0.98 & 0.6\\
	s010 & 00:46:00.67 & +61:44:14.5 & 11.0 & 1.41 & -5.2 & -0.22 & 0.98 & 0.4\\
	s011 & 00:42:46.90 & +61:36:23.0 & 11.2 & 1.43 & -5.27 & -0.14 & 0.98 & 0.8\\
	s012 & 00:43:07.74 & +61:29:52.3 & 11.3 & 1.48 & -5.36 & -0.2 & 0.98 & 0.6\\
	s013 & 00:44:12.81 & +61:51:01.8 & 11.4 & 1.3 & -5.91 & -0.46 & 0.85 & -\\
	s014 & 00:40:12.49 & +61:41:30.0 & 11.4 & 1.49 & -3.95 & -0.77 & 0.68 & -\\
	s015 & 00:43:25.62 & +61:48:51.6 & 11.5 & 1.48 & -5.26 & -0.08 & 0.98 & 0.7\\
	s016 & 00:43:36.93 & +61:53:40.1 & 12.0 & 1.45 & -6.04 & 0.03 & 0.95 & 0.6\\
	s017 & 00:46:41.32 & +61:44:27.4 & 12.1 & 1.5 & -5.34 & -0.06 & 0.97 & 0.4\\
	s018 & 00:43:31.00 & +61:48:10.1 & 12.1 & 1.37 & -4.77 & -0.82 & 0.85 & 0.1\\
	s019 & 00:41:02.54 & +61:33:51.5 & 12.3 & 1.39 & -5.32 & -0.08 & 0.98 & 0.8\\
	s020 & 00:44:16.53 & +61:50:44.0 & 12.3 & 1.42 & -5.29 & -0.15 & 0.98 & 0.7\\
	s021 & 00:40:30.79 & +61:40:17.5 & 12.5 & 1.15 & -5.32 & 0.09 & 0.64 & -\\
	s022 & 00:44:11.73 & +61:40:16.6 & 12.6 & 1.44 & -5.42 & -0.09 & 0.98 & 0.8\\
	s023 & 00:41:35.10 & +61:39:42.2 & 12.6 & 1.45 & -5.35 & -0.04 & 0.98 & 0.8\\
	s024 & 00:44:20.73 & +61:49:45.1 & 13.0 & 1.36 & -5.35 & -0.03 & 0.97 & 1.0\\
	s025 & 00:45:25.60 & +61:40:23.2 & 13.1 & 1.4 & -4.74 & -0.15 & 0.96 & 0.1\\
	s026 & 00:43:18.87 & +61:41:14.4 & 13.2 & 1.45 & -5.37 & -0.19 & 0.98 & 0.9\\
	s027 & 00:44:47.94 & +61:53:38.3 & 13.3 & 1.44 & -5.39 & -0.06 & 0.98 & 0.9\\
	s028 & 00:43:05.53 & +61:53:43.8 & 13.3 & 1.43 & -5.31 & -0.16 & 0.98 & 0.7\\
	s029 & 00:42:50.93 & +61:43:21.8 & 13.4 & 1.45 & -5.42 & -0.25 & 0.98 & 0.9\\
	s030 & 00:43:25.72 & +61:40:31.8 & 13.4 & 1.41 & -5.97 & -0.14 & 0.97 & 0.5\\
	s031 & 00:40:52.30 & +61:43:14.4 & 13.6 & 1.48 & -4.64 & -0.22 & 0.93 & 0.1\\
	s032 & 00:44:14.76 & +61:40:21.7 & 13.9 & 1.42 & -5.22 & -0.04 & 0.98 & 0.7\\
	s033 & 00:41:14.85 & +61:35:30.2 & 14.0 & 1.25 & -5.67 & -1.32 & 0.52 & -\\
	s034 & 00:43:48.95 & +61:52:51.8 & 14.1 & 1.4 & -5.38 & -0.35 & 0.98 & 0.5\\
	s035 & 00:45:35.25 & +61:55:03.4 & 14.1 & 1.45 & -5.28 & -0.23 & 0.98 & 0.7\\
	s036 & 00:43:43.35 & +61:43:04.5 & 14.2 & 1.39 & -5.84 & -0.41 & 0.96 & 0.3\\
	s037 & 00:45:06.51 & +61:51:52.7 & 14.2 & 1.4 & -5.18 & -0.46 & 0.97 & 0.3\\
	s038 & 00:45:19.99 & +61:42:11.4 & 14.5 & 1.58 & -5.4 & -0.11 & 0.88 & -\\
	s039 & 00:45:18.66 & +61:55:22.1 & 14.6 & 1.52 & -5.66 & -0.15 & 0.96 & 0.5\\
	s040 & 00:41:08.21 & +61:49:00.5 & 14.7 & 1.43 & -5.33 & 0.07 & 0.98 & 0.8\\
	s041 & 00:40:37.06 & +61:53:53.8 & 14.9 & 1.43 & -5.53 & -0.13 & 0.98 & 0.8\\
	s042 & 00:40:12.88 & +61:41:31.4 & 14.9 & 1.49 & -5.49 & 0.15 & 0.97 & -\\
	s043 & 00:45:49.65 & +61:42:43.5 & 15.1 & 1.39 & -5.48 & 0.05 & 0.98 & 0.8\\
	s044 & 00:46:07.88 & +61:45:06.1 & 15.1 & 1.48 & -6.12 & 0.3 & 0.89 & 0.5\\
	s045 & 00:40:47.91 & +61:46:33.6 & 15.2 & 1.46 & -5.37 & -0.2 & 0.98 & 0.7\\
	s046 & 00:45:09.60 & +61:51:31.3 & 15.2 & 1.38 & -5.44 & 0.1 & 0.98 & 0.8\\
	s047 & 00:45:36.42 & +61:57:58.9 & 15.5 & 1.42 & -6.39 & -0.22 & 0.89 & -\\
	s048 & 00:45:08.93 & +61:38:36.1 & 15.5 & 1.48 & -5.46 & 0.11 & 0.97 & 0.6\\
	s049 & 00:46:25.80 & +61:53:37.6 & 15.6 & 1.43 & -3.77 & -0.88 & 0.64 & -\\
	s050 & 00:40:27.47 & +61:46:21.9 & 15.6 & 1.42 & -5.44 & -0.04 & 0.98 & 0.8\\
	*s051 & 00:44:35.54 & +61:25:28.8 & 15.6 & 1.33 & -3.44 & -0.44 & 0.62 & -\\
	s052 & 00:44:46.99 & +61:42:53.8 & 15.7 & 1.34 & -3.5 & -0.37 & 0.65 & -\\
	s053 & 00:41:54.71 & +61:44:10.5 & 15.7 & 1.46 & -5.26 & 0.03 & 0.98 & 0.7\\
	s054 & 00:40:37.32 & +61:57:15.7 & 15.8 & 1.47 & -5.39 & -0.06 & 0.98 & 0.7\\
	s055 & 00:42:49.15 & +62:07:45.6 & 15.9 & 1.35 & -7.13 & 0.78 & 0.51 & -\\
	s056 & 00:41:41.22 & +61:53:59.0 & 15.9 & 1.36 & -4.76 & -0.33 & 0.95 & 0.1\\
	s057 & 00:43:35.36 & +61:44:31.5 & 16.0 & 1.4 & -5.35 & -0.13 & 0.98 & 0.8\\
	s058 & 00:42:40.78 & +61:37:22.3 & 16.0 & 1.38 & -5.36 & -0.16 & 0.98 & 0.9\\
	s059 & 00:40:42.27 & +61:36:09.4 & 16.0 & 1.44 & -5.23 & -0.07 & 0.98 & 0.5\\
	s060 & 00:43:16.77 & +61:50:15.2 & 16.1 & 1.43 & -5.82 & 1.22 & 0.55 & -\\
	s061 & 00:44:07.20 & +62:03:30.4 & 16.2 & 1.41 & -5.22 & -0.15 & 0.98 & 0.6\\
	s062 & 00:42:30.29 & +61:40:47.7 & 16.3 & 1.4 & -5.53 & -0.2 & 0.98 & 0.8\\
	s063 & 00:43:14.64 & +62:09:21.2 & 16.3 & 1.24 & -4.44 & 0.14 & 0.72 & -\\

		\hline

	\end{tabular}
\end{table*}

\begin{table*}
	\centering
	\contcaption{List of cluster members with membership probability > 50 per\,cent. $G_{Gaia}$ magnitude, proper motion and parallaxes are from Gaia DR2. Symbol "*" refers to rejected star.}
	\label{tab:List50_1}
	\begin{tabular}{lcccccccc} 
		\hline
	Star & Right Ascention, & Declination, & $G_{Gaia},$& Parallax, & $ \mu_{\alpha}~cos\delta$, & $\mu_{\delta}$,  & Probability & Probability from\\
	& h:m:s           & d:m:s       & mag & mas & mas\,year$^{-1}$ & mas\,year$^{-1}$ & & \citet{CantatGaudin}    \\
	\hline
	s064 & 00:40:31.44 & +61:56:55.0 & 16.3 & 1.33 & -5.71 & -0.06 & 0.96 & -\\
	s065 & 00:42:21.24 & +61:39:43.0 & 16.3 & 1.42 & -5.32 & -0.12 & 0.98 & 0.7\\
	s066 & 00:41:15.67 & +61:31:05.9 & 16.5 & 1.56 & -5.34 & 0.03 & 0.91 & -\\
	s067 & 00:43:15.29 & +62:07:26.1 & 16.5 & 1.34 & -5.59 & 0.22 & 0.95 & 0.5\\
	s068 & 00:43:55.29 & +61:52:38.6 & 16.5 & 1.57 & -5.14 & 0.04 & 0.9 & -\\
	s069 & 00:44:09.51 & +61:42:54.9 & 16.7 & 1.35 & -5.18 & -0.18 & 0.97 & 0.4\\
	s070 & 00:44:05.50 & +62:02:13.8 & 16.8 & 1.43 & -5.65 & 0.05 & 0.98 & 0.7\\
	s071 & 00:42:41.18 & +61:49:10.1 & 16.8 & 1.45 & -5.73 & 1.07 & 0.65 & -\\
	s072 & 00:44:24.93 & +61:49:58.5 & 16.9 & 1.51 & -5.3 & 0.25 & 0.95 & 0.5\\
	s073 & 00:44:17.30 & +61:55:07.7 & 16.9 & 1.4 & -5.4 & -0.15 & 0.98 & 0.7\\
	s074 & 00:41:50.85 & +61:58:22.4 & 17.0 & 1.43 & -5.27 & -0.04 & 0.98 & 0.4\\
	s075 & 00:44:10.57 & +62:02:03.7 & 17.0 & 1.36 & -4.01 & -1.01 & 0.66 & -\\
	s076 & 00:45:06.70 & +61:37:39.5 & 17.2 & 1.38 & -5.77 & -0.19 & 0.97 & 0.4\\
	s077 & 00:43:47.32 & +62:05:31.0 & 17.2 & 1.7 & -4.31 & -0.01 & 0.5 & -\\
	s078 & 00:42:31.63 & +61:45:30.6 & 17.2 & 1.42 & -5.5 & -0.18 & 0.98 & 0.6\\
	s079 & 00:42:30.56 & +61:32:27.6 & 17.2 & 1.55 & -5.49 & -0.32 & 0.92 & -\\
	s080 & 00:43:03.84 & +61:45:32.4 & 17.3 & 1.45 & -6.0 & 0.26 & 0.94 & 0.6\\
	s081 & 00:43:05.72 & +61:39:49.6 & 17.4 & 1.17 & -6.53 & 0.01 & 0.55 & -\\
	s082 & 00:43:28.66 & +61:49:18.5 & 17.4 & 1.38 & -5.2 & -0.1 & 0.98 & 0.5\\
	s083 & 00:45:42.85 & +61:38:36.2 & 17.4 & 1.36 & -5.01 & 0.6 & 0.81 & 0.1\\
	s084 & 00:43:16.66 & +61:49:09.3 & 17.5 & 1.54 & -5.11 & 0.11 & 0.93 & -\\
	s085 & 00:46:15.54 & +61:41:18.6 & 17.5 & 1.51 & -5.12 & 0.16 & 0.96 & 0.4\\
	s086 & 00:40:46.66 & +61:43:58.5 & 17.5 & 1.27 & -5.16 & -0.2 & 0.89 & -\\
	s087 & 00:41:22.97 & +61:47:53.4 & 17.6 & 1.38 & -5.27 & -0.06 & 0.98 & 0.6\\
	s088 & 00:44:44.44 & +61:42:58.5 & 17.6 & 1.5 & -5.22 & -0.34 & 0.97 & 0.5\\
	s089 & 00:40:16.25 & +61:50:39.7 & 17.6 & 1.46 & -5.42 & -0.38 & 0.98 & -\\
	s090 & 00:43:53.48 & +61:43:42.8 & 17.7 & 1.44 & -5.44 & 0.39 & 0.95 & 0.4\\
	s091 & 00:42:57.00 & +61:54:40.7 & 17.8 & 1.54 & -5.71 & 0.93 & 0.62 & -\\
	s092 & 00:42:45.07 & +61:48:03.7 & 17.8 & 1.36 & -4.84 & -0.56 & 0.92 & 0.2\\
	s093 & 00:43:25.87 & +61:39:56.7 & 17.8 & 1.59 & -5.33 & 0.05 & 0.88 & -\\
	s094 & 00:44:49.01 & +61:47:40.1 & 17.8 & 1.3 & -5.52 & 0.07 & 0.93 & -\\
	s095 & 00:41:02.70 & +61:31:45.3 & 17.9 & 1.32 & -4.97 & -0.42 & 0.91 & -\\
	s096 & 00:44:10.38 & +61:47:25.8 & 17.9 & 1.51 & -5.47 & -0.27 & 0.97 & 0.5\\
	s097 & 00:44:54.76 & +61:40:56.6 & 17.9 & 1.17 & -5.31 & -0.44 & 0.67 & -\\
	s098 & 00:42:45.38 & +61:46:45.1 & 18.0 & 1.39 & -5.51 & 0.37 & 0.95 & 0.4\\
	s099 & 00:42:26.85 & +61:54:00.4 & 18.0 & 1.22 & -5.26 & -0.31 & 0.77 & -\\
	s100 & 00:40:56.82 & +61:57:39.0 & 18.0 & 1.35 & -5.44 & -0.12 & 0.97 & -\\
	s101 & 00:42:11.84 & +61:34:43.9 & 18.1 & 1.65 & -4.44 & -0.69 & 0.56 & -\\
	s102 & 00:42:08.48 & +61:55:57.5 & 18.1 & 1.55 & -5.43 & 0.36 & 0.88 & -\\
	s103 & 00:41:34.23 & +61:29:19.6 & 18.2 & 1.41 & -5.27 & -0.31 & 0.98 & -\\
	s104 & 00:46:24.32 & +61:53:44.3 & 18.2 & 1.33 & -4.87 & -1.32 & 0.62 & -\\
	s105 & 00:43:28.66 & +61:47:18.7 & 18.3 & 1.44 & -4.8 & -0.86 & 0.85 & -\\
	s106 & 00:44:04.17 & +61:53:30.2 & 18.4 & 1.46 & -5.92 & 0.05 & 0.96 & -\\
	s107 & 00:46:55.77 & +61:43:46.2 & 18.4 & 1.49 & -5.2 & -0.48 & 0.96 & -\\
	s108 & 00:44:17.22 & +61:50:32.7 & 18.4 & 1.21 & -5.33 & -0.31 & 0.77 & -\\
	s109 & 00:44:17.00 & +61:43:21.0 & 18.4 & 1.64 & -5.34 & 0.36 & 0.69 & -\\
	s110 & 00:42:50.68 & +61:48:46.0 & 18.5 & 1.49 & -5.11 & -0.12 & 0.97 & -\\
	s111 & 00:42:13.88 & +61:57:02.0 & 18.5 & 1.13 & -5.27 & -0.31 & 0.6 & -\\
	s112 & 00:46:16.24 & +61:32:28.7 & 18.5 & 1.49 & -5.08 & -0.38 & 0.97 & -\\
	s113 & 00:42:13.54 & +61:59:10.7 & 18.5 & 1.25 & -5.61 & 0.34 & 0.8 & -\\
	s114 & 00:44:46.60 & +61:44:02.6 & 18.6 & 1.59 & -5.7 & -0.06 & 0.86 & -\\
	s115 & 00:44:26.51 & +61:47:43.8 & 18.6 & 1.72 & -5.02 & -0.36 & 0.53 & -\\
	s116 & 00:43:46.04 & +61:58:12.1 & 18.6 & 1.22 & -5.86 & -0.68 & 0.68 & -\\
	s117 & 00:43:34.31 & +62:11:00.3 & 18.6 & 1.44 & -5.19 & -0.38 & 0.98 & -\\
	s118 & 00:47:04.85 & +61:41:07.8 & 18.6 & 1.28 & -4.13 & -0.51 & 0.73 & -\\
	s119 & 00:43:47.53 & +61:53:27.5 & 18.6 & 1.36 & -4.99 & 0.01 & 0.97 & -\\
	s120 & 00:44:20.64 & +61:45:45.6 & 18.7 & 1.33 & -5.2 & 0.05 & 0.96 & -\\
	s121 & 00:44:30.13 & +61:46:20.2 & 18.7 & 1.17 & -5.94 & -0.11 & 0.66 & -\\
	s122 & 00:41:25.27 & +61:37:41.5 & 18.7 & 1.32 & -5.01 & -0.78 & 0.83 & -\\
	s123 & 00:45:09.05 & +61:41:49.0 & 18.8 & 1.44 & -5.51 & -0.32 & 0.98 & -\\
	s124 & 00:44:18.79 & +62:03:44.7 & 18.9 & 1.25 & -4.97 & -0.88 & 0.71 & -\\
	s125 & 00:46:21.85 & +61:48:37.8 & 19.0 & 1.44 & -5.32 & 0.18 & 0.97 & -\\
	s126 & 00:45:45.89 & +61:29:46.5 & 19.1 & 1.68 & -5.81 & -0.24 & 0.61 & -\\
	s127 & 00:44:48.58 & +61:34:29.6 & 19.2 & 1.35 & -5.43 & -0.48 & 0.95 & -\\
	s128 & 00:44:54.27 & +61:50:36.5 & 19.2 & 1.52 & -5.29 & -0.46 & 0.94 & -\\
	s129 & 00:42:49.29 & +61:59:25.9 & 19.3 & 1.68 & -6.1 & -0.66 & 0.56 & -\\
		
		\hline

	\end{tabular}
\end{table*}


\bsp	
\label{lastpage}
\end{document}